\documentclass{amsart}

\input{epsf}

\newtheorem{thm}{Theorem}[section]
\newtheorem{cor}[thm]{Corollary}
\newtheorem{lem}[thm]{Lemma}

\theoremstyle{definition}

\theoremstyle{remark}

\numberwithin{equation}{section}

\newcommand{\bs}{\medskip}

\begin{document}

\title[Krein's Formula And Heat-Kernel Expansion ...]
{Krein's Formula And Heat-Kernel Expansion For Some Differential
Operators With A Regular Singularity}%
\author
{H.\ Falomir and P.A.G.\ Pisani}

\address{IFLP (CONICET) and Departamento de F{\'\i}sica - Facultad de Ciencias
Exactas, Universidad Nacional de La Plata, C.C. 67 (1900) La
Plata, Argentina}



\begin{abstract}
We get a generalization of Krein's formula -which relates the
resolvents of different selfadjoint extensions of a differential
operator with regular coefficients- to the non-regular case
$A=-\partial_x^2+(\nu^2-1/4)/x^2+V(x)$, where $0<\nu<1$ and $V(x)$
is an analytic function of $x\in\mathbb{R}^+$ bounded from below.
We show that the trace of the heat-kernel $e^{-tA}$ admits a
non-standard small-t asymptotic expansion which contains, in
general, integer powers of $t^\nu$. In particular, these powers
are present for those selfadjoint extensions of $A$ which are
characterized by boundary conditions that break the local formal
scale invariance at the singularity.
\end{abstract}

\maketitle
\section{Introduction}

In Quantum Field Theory the effective action, the free energy and
other physical quantities related to the vacuum state are
generically divergent and require a renormalization procedure. A
powerful and elegant regularization scheme relies on the small-$t$
asymptotic expansion of the trace of the heat-kernel $e^{-t A}$
corresponding to an elliptic differential operator $A$ determined
by the Lagrangian of the theory  (see, e.g.,
\cite{Elizalde,KK,DV}.)

\bs

It is well-known \cite{Seeley} that for an elliptic boundary value
problem in an $m$-dimensional compact manifold with boundary,
described by a differential operator $A$ of order $d$, with smooth
coefficients and defined on a domain of functions subject to local
boundary conditions, the heat-kernel trace admits a small-$t$
asymptotic expansion given by
\begin{equation}\label{heat-trace}
  {\rm Tr}\{e^{-t A}\}\sim \sum_{n=0}^\infty a_n(A)\, t^{(n-m)/d}\,,
\end{equation}
where the coefficients $a_n(A)$ are integrals on the manifold and
its boundary of geometrical invariants \cite{Gilkey}.

\bs

However, not much is known about the heat-kernel trace asymptotic
expansion for the case of differential operators with singular
coefficients (see chapter $6$ of \cite{DV}.)

\bs

In 1980 C.J.\ Callias and C.H.\ Taubes \cite{Callias1p} have
pointed out that for some differential operators with singular
coefficients the heat-kernel trace asymptotic expansion in terms
of powers of the form $t^{(n-m)/d}$ is ill-defined and conjectured
that more general powers of $t$, as well as $\log t$ terms, could
appear.

\bs

In the present article we will study the small-$t$ asymptotic
expansion of the heat-kernel trace of the differential operator
\begin{equation}\label{mod}
    A=-\partial_x^2+\frac{\nu^2-1/4}{x^2}+V(x)\,,
\end{equation}
where $\nu\in(0,1)\subset\mathbb{R}$ and $V(x)$ is an analytic
function of $x\in\mathbb{R}^+$.

\bs

The operator (\ref{mod}) defined on
$\mathcal{D}(A):=\mathcal{C}_0^\infty(\mathbb{R}^+)$, the set of
smooth functions on $\mathbb{R}^+$ with compact support out of the
origin, admits a one-parameter family of selfadjoint extensions,
$A^\theta$ with $\theta\in\mathbb{R}$. These selfadjoint
extensions are characterized by the {\em boundary condition} the
functions in their domains satisfy at the singular point $x=0$.
They describe a different physical system, and its spectral
properties depend on the different  behavior of the functions at
the singularity.

\bs

Since the heat-kernel $e^{-tA^\theta}$ corresponding to an
arbitrary selfadjoint extension $A^\theta$ is not trace-class, we
will study the trace of the difference
$e^{-tA^\theta}-e^{-tA^\infty}$, where $A^\infty$ denotes the
Friedrichs extension. We will show in Section \ref{nocom} that
this trace admits an asymptotic expansion of the form
\begin{equation}\label{weshow}
    {\rm Tr}\left\{e^{-tA^\theta}-e^{-tA^\infty}\right\}\sim
    \sum_{n=0}^\infty a_n(\nu,V)\,t^{\frac n 2}
    +\sum_{N,n=1}^\infty b_{N,n}(\nu,V)\,\theta^N\,
    t^{\nu N+\frac n 2-\frac 1 2}\,.
\end{equation}
The coefficients $a_n(\nu,V),b_{N,n}(\nu,V)$ can be recursively
computed for each potential $V(x)$. Notice that the singular term
in (\ref{mod}) not only contributes to the coefficients
$a_n(\nu,V)$ of the standard terms but also leads to the presence
of powers of $t$ whose exponents are not half-integers but depend
on the ``external'' parameter $\nu$.

\bs

It is also remarkable that these terms are absent only for
$\theta=0$ and $\theta=\infty$, which correspond to the
selfadjoint extensions characterized by scale invariant boundary
conditions at the singular point $x=0$. Indeed, the first two
terms in the {\small R.H.S.\ }of eq.\ (\ref{mod}), which are
dominant for $x\simeq 0$, present the same scaling dimension.
However, the behavior of the wave functions at the origin breaks,
in general, this local scale invariance. Only for the selfadjoint
extensions characterized by $\theta=0$ or $\theta=\infty$ is the
boundary condition also scale invariant (see eq.\
(\ref{comenelorieq}).)

\bs

There is a dimensional argument in favor of the plausibility of
the result in (\ref{weshow}). Notice that the parameter $\theta$
in (\ref{comenelorieq}) has dimensions $[{\rm length}]^{-2\nu}$
and, due to the analyticity of $V(x)$, the dimensions of any other
parameter in the problem is an integer power of the length. Since
$t$ has dimensions $[{\rm length}]^{2}$, if the coefficients of
the asymptotic expansion of the heat-kernel trace were to depend
on the boundary conditions by means of a polynomial dependence on
$\theta$, then this expansion should contain integer powers of
$t^\nu$. Consequently, the only selfadjoint extensions for which
these powers are to be absent are $\theta=0$ and $\theta=\infty$.

\bs

As a matter of fact, the ``functorial method'' \cite{G,KK,DV}
which has been widely used to determine the coefficients of the
heat-kernel expansion in the regular case can be also applied to
operator (\ref{mod}) to determine some of the $a_n(\nu,V)$ and
$b_{N,n}(\nu,V)$ in expression (\ref{weshow}). The asymptotic
expansion (\ref{weshow}) generalizes the results in \cite{FPW,FP2}
where some singular Schr\"odinger and Dirac operators were
considered.

\bs

There is a second new result in the present article which lead us
to the derivation of the asymptotic expansion (\ref{weshow}). By
generalizing Krein's formula \cite{Krein's1} (see also \cite{A-G})
we find a relation between the resolvents corresponding to
different selfadjoint extensions of operator (\ref{mod}). Since
there exist two selfadjoint extensions, namely $A^0$ and
$A^\infty$, for which the $\nu$ dependent exponents of $t$ in
(\ref{weshow}) are absent, expansion (\ref{weshow}) will come out
as a consequence of this relation.

\medskip

Schr\oe dinger operators defined by a singular potential whose
leading behavior near the singularity is given by (\ref{mod}) have
been studied as models of conformal invariance in quantum
mechanics \cite{DA-F-F}, in Calogero models \cite{Calogero}, in
SUSY breaking in quantum mechanics \cite{FP2} and in cosmic
strings \cite{Vilenkin:1984ib}. Since the dynamics of quantum
fields on black holes' backgrounds is described by operators
similar to (\ref{mod}), there exists a microscopic description of
black holes in the vicinity of the horizon in terms of conformal
models \cite{Gibbons}. In this context, in which the operator
(\ref{mod}) is relevant, particular attention to the most general
boundary conditions has been given in \cite{Birmingham}.

\bs

Differential operators with a singular coefficient given by
(\ref{mod}) are also obtained from the Laplacian on manifolds with
conical singularities where the parameter $\nu$ is related to the
deficiency angle. The asymptotic expansion of the heat-kernel of
the Laplacian on manifolds with conical singularities has been
considered, probably for the first time, in \cite{carcone}. This
problem has been also studied in \cite{brusee}; however, the most
general boundary conditions at the singularity where not
considered there.

\bs

More recently, E.\ Mooers \cite{Mooers} studied the selfadjoint
extensions of the Laplacian acting on differential forms on a
manifold with a conical singularity and showed that the asymptotic
expansion of the heat-kernel trace contains powers of $t$ whose
exponents depend on the deficiency angle of the
singularity\,\footnote{Although this result is confirmed by our
calculations we obtain a different value for the corresponding
coefficient (see page (4) of \cite{Mooers}.)}.

\bs

In Section \ref{SAE} we describe the selfadjoint extensions of
operator (\ref{mod}) and in Section \ref{singu} we generalize
Krein's formula to this type of singular operators. Finally, in
Section \ref{nocom}, we use this generalization to establish
expansion (\ref{weshow}) that describes the small-$t$ asymptotic
expansion of the heat-kernel trace.

\section{Self-adjoint Extensions}\label{SAE}

Let us consider the one-dimensional differential operator $A$
given by (\ref{mod}) defined on
$\mathcal{D}(A):=\mathcal{C}^\infty_0(\mathbb{R^+})\subset\mathbf{L}_2(\mathbb{R^+})$.
First, notice that the first two terms in the {\small R.H.S.\ }of
expression (\ref{mod}) have the same scaling properties at the
singular point $x=0$. This fact will be essential in the
following.

Next, we describe the behavior at the singular point $x=0$ of the
functions in $\mathcal{D}(A^\dagger)$.

{\thm
\begin{equation}\label{comenelorieq}
    \psi\in\mathcal{D}(A^{\dagger})\rightarrow\psi(x)=C[\psi]\,\left(
    x^{-\nu+1/2}+
    \theta_{\psi}\,x^{\nu+1/2}\right)+O(x^{3/2})\,,
\end{equation}
for $x\rightarrow 0^+$ and some constants
$C[\psi],\theta_\psi\in\mathbb{C}$.}\label{comenelori}

\bs

{\noindent\bf Proof:} See the
Appendix.\begin{flushright}$\Box$\end{flushright}

{\cor \label{Corooo}
\begin{equation}\label{corolario}
    \phi,\psi\in\mathcal{D}(A^{\dagger})\rightarrow
    (\phi,A^{\dagger}\psi)-(A^{\dagger}\phi,\psi)=
    C^*[\phi]C[\psi]\left(
    \theta^*_{\phi}-\theta_{\psi}\right)\,.
\end{equation}}

{\noindent\bf Remark:} By choosing $\psi=\phi$ we conclude that
for all $\psi\in\mathcal{D}(A^\dagger)$ the parameter
$\theta_\psi$ defined by Theorem $2.1$ is real.

\bs

{\noindent\bf Proof:} Expression (\ref{corolario}) follows from an
integration by parts in its {\small L.H.S.\ }using Theorem
$2.1$.\begin{flushright}$\Box$\end{flushright}

\bs

As a consequence of Corollary \ref{Corooo} the selfadjoint
extensions $A^{\theta}$ of the differential operator $A$ are
characterized by a real parameter $\theta$, being their domains
defined by
\begin{equation}\label{saesing}
    \mathcal{D}(A^{\theta}):=
    \left\{\phi\in\mathcal{D}(A^{\dagger}):\theta_{\phi}=
    \theta\right\}\,,
\end{equation}
where $\theta_\phi$ is defined according to Theorem
\ref{comenelori}. The parameter $\theta$ thus determine the
boundary condition at the singularity.

\bs

There exists another selfadjoint extension (see the Appendix),
which we denote by $A^{\infty}$, whose domain is given by,
\begin{equation}\label{beta}
    \mathcal{D}(A^{\infty})=
    \left\{\phi\in\mathcal{D}(A^{\dagger}):
    \phi(x)=C[\phi]\,x^{\nu+1/2}+O(x^{3/2})\,,\ {\rm with\ }C[\phi]\in
    \mathbb{C}\right\}\,.
\end{equation}

\section{Generalization of Krein's formula.}\label{singu}

The non-regular differential operator $A$, given by expression
(\ref{mod}), defined on
$\mathcal{D}(A):=\mathcal{C}_0^\infty(\mathbb{R}^+)$ admits an
infinite family of selfadjoint extensions $A^\theta$ characterized
by a real parameter $\theta$. As we have shown, this parameter
describes the boundary condition at the singularity. The purpose
of this Secction is to establish a relation between the resolvents
corresponding to these selfadjoint extensions. We will
consequently obtain a generalization of Krein's formula.

\bs

The kernel $G_{\theta}(x,x',\lambda)$ of the resolvent
$(A^{\theta}-\lambda)^{-1}$ can be written as
\begin{equation}\label{solres}
    G_\theta(x,x',\lambda)=-\frac{1}{W(\lambda)}\left\{\
    \Theta(x'-x)L_\theta(x,\lambda)R(x',\lambda)+
    \Theta(x-x')L_\theta(x',\lambda)R(x,\lambda)\right\}\,,
\end{equation}
where $\Theta(x)$ is Heaviside function and
$L_\theta(x,\lambda),R(x,\lambda)\in{\rm Ker}(A^\dagger-\lambda)$.
The latter is square integrable at $x\rightarrow \infty$ and the
former satisfies the boundary condition
\begin{equation}\label{latac}
    L_\theta(x,\lambda)=x^{-\nu+1/2}+\theta\,x^{\nu+1/2}+O(x^{3/2})\,,
\end{equation}
at $x\rightarrow 0^+$. $W(\lambda)$ is their Wronskian, which is
independent of $x$.

\bs

As a first step, we will find a relation between the resolvents
corresponding to $\theta=\infty$ and $\theta=0$. In order to do
this, we consider the equation,
\begin{equation}\label{pro}
    (A^\theta-\lambda)\phi^\theta(x,\lambda)=f(x)\,
\end{equation}
whose solutions for $\theta=0$ and $\theta=\infty$ are given by
\begin{eqnarray}
    \ \ \ \phi^{\infty}(x,\lambda)=
    \int_0^{\infty}G_{\infty}(x,x',\lambda)f(x')\,dx'=
    \phi^{\infty}(\lambda)\,x^{\nu+1/2}+O(x^{3/2})\,,\label{fii}\\
    \phi^{0}(x,\lambda)=
    \int_0^{\infty}G_{0}(x,x',\lambda)f(x')\,dx'=
    \phi^{0}(\lambda)\,x^{-\nu+1/2}+O(x^{3/2})\,.\label{fi0}
\end{eqnarray}
Notice that
\begin{equation}
    \phi^{\infty}(\lambda)=
    \int_0^{\infty}G_{\infty}(x',\lambda)f(x')\,dx'\,,\label{fiienelori}\qquad
    \phi^{0}(\lambda)=
    \int_0^{\infty}G_{0}(x',\lambda)f(x')\,dx'\, , 
\end{equation}
where
\begin{equation}
    G_{\infty}(x',\lambda):=\lim_{x\rightarrow
    0}x^{-\nu-1/2}\,G_{\infty}(x,x',\lambda)\,,\label{Rxp}\qquad
    G_{0}(x',\lambda):=\lim_{x\rightarrow
    0}x^{\nu-1/2}\,G_{0}(x,x',\lambda)\,.
\end{equation}
\begin{lem}\label{lemcon}
\begin{equation}\label{0vsinf}
    \phi^{0}(x,\lambda)=\phi^{\infty}(x,\lambda)+2\nu\, G_{\infty}
    (x,\lambda)\,\phi^{0}(\lambda)\,.
\end{equation}
\end{lem}
{\noindent\bf Proof:}
\begin{eqnarray}
\phi^{0}(x,\lambda)-\phi^{\infty}(x,\lambda)
=\int_0^\infty\left[G_{0}(x,x',\lambda)-
G_{\infty}(x,x',\lambda)\right](A^0-\lambda)\phi^0(x',\lambda)\,dx'=\nonumber\\
=\mbox{}-\lim_{x'\rightarrow 0^+}\{\left[G_{0}(x,x',\lambda)-
G_{\infty}(x,x',\lambda)\right]\partial_{x'}\phi^0(x',\lambda)-
\nonumber\\
\mbox{}-\partial_{x'}\left[G_{0}(x,x',\lambda)-
G_{\infty}(x,x',\lambda)\right]\phi^0(x',\lambda)\}= 2\nu\,
G_{\infty}(x,\lambda)\,\phi^{0}(\lambda)\,.\nonumber
\end{eqnarray}
\begin{flushright}$\Box$\end{flushright}
Taking the limit $x\rightarrow 0^+$ in equation (\ref{0vsinf}) we
obtain
\begin{equation}\label{Renelori}
    G_{\infty}(x,\lambda)=\frac{1}{2\nu}
    \left(x^{-\nu+1/2}-K(\lambda)^{-1}x^{\nu+1/2}\right)+O(x^{3/2})\,,
\end{equation}
where
\begin{equation}\label{Renelori2}
    K(\lambda):=\frac{\phi^{0}(\lambda)}{\phi^{\infty}(\lambda)}\,.
\end{equation}
Notice that $K(\lambda)$ can be computed by studying the behavior
at the singularity of the kernel of the resolvent corresponding to
the extension $\theta=\infty$.

\bs

Replacing $\phi^{0}(\lambda)$ from eq.\ (\ref{Renelori2}) into
eq.\ (\ref{0vsinf}) one can express the solution
$\phi^{0}(x,\lambda)$ corresponding to $\theta=0$ by means of data
related to the selfadjoint extension corresponding to
$\theta=\infty$,
\begin{equation}\label{comp}
    \phi^{0}(x,\lambda)=\phi^{\infty}(x,\lambda)+2\nu K(\lambda)\,
    G_{\infty}(x,\lambda)\,\phi^{\infty}(\lambda)\,.
\end{equation}
Next, we will establish a similar expression giving the resolvent
for an arbitrary selfadjoint extension in terms of data related to
the boundary conditions corresponding to $\theta=\infty$.
\begin{lem}\label{rem}
\begin{equation}\label{fithe}
    \phi^{\theta}(x,\lambda)=\phi^{\infty}(x,\lambda)+2\nu
 \left(K(\lambda)^{-1}+\theta\right)^{-1}G_{\infty}(x,\lambda)\,
 \phi^{\infty}(\lambda)\,.
\end{equation}
\end{lem}

\noindent{\bf Proof:} Eq.\ (\ref{0vsinf}) shows that the
difference between both sides of expression (\ref{fithe}) belongs
to ${\rm Ker}(A^\dagger-\lambda)$. Moreover, eqs.\ (\ref{fii}) and
(\ref{Renelori}) show that both sides of (\ref{fithe}) belong to
$\mathcal{D}(A^\theta)$. The proof follows by virtue of the
uniqueness of the solution of equation
(\ref{pro}).\begin{flushright}$\Box$\end{flushright}

Finally, from eqs.\ (\ref{comp}) and (\ref{fithe}) it is
straightforward to obtain the following theorem:
\begin{thm}[Generalization of Krein's formula]\label{elthm}
\begin{equation}\label{kreinsing}
    \left(A^{\theta}-\lambda\right)^{-1}-
    \left(A^{\infty}-\lambda\right)^{-1}=
    \frac{\left(A^{0}-\lambda\right)^{-1}-
    \left(A^{\infty}-\lambda\right)^{-1}}{1+\theta\, K(\lambda)}\,.
\end{equation}
\end{thm}

\bs

In the next section we will show that the asymptotic expansion of
$K(\lambda)$ for large $|\lambda|$ presents powers of $\lambda$
whose exponents depend on the parameter $\nu$. This leads, due to
the relation between the resolvent and the heat-kernel, to the
asymptotic series (\ref{weshow}).

\section{Asymptotic expansion of the resolvent}\label{nocom}

In this section we will show that $K(\lambda)$ admits a
large-$|\lambda|$ asymptotic expansion in powers of $\lambda$ with
$\nu$-dependent exponents. Since we can consider $\lambda$ in the
negative real semi-axis of the complex plane, we will study the
solutions $\psi$ of equation
\begin{equation}\label{asiecu}
    (A+z)\psi(x,z)=0\,,
\end{equation}
for large $z\in\mathbb{R}^+$. In particular, due to eqs.\
(\ref{Renelori}) and (\ref{kreinsing}) we just need to consider
solutions satisfying the boundary conditions corresponding to
$\theta=\infty$ and $\theta=0$.

\bs

Taking into account the scaling properties of the first two terms
in (\ref{mod}) it is convenient to define a new variable
$y:=\sqrt{z}\,x\in\mathbb{R}^+$. Equation (\ref{asiecu}) can then
be written as
\begin{equation}\label{asiecucamvar}
    \left(-\partial^2_y+\frac{\nu^2-1/4}{y^2}+1+\frac 1 z
    \,V(y/\sqrt{z})\right)\psi(y/\sqrt{z},z)=0\,.
\end{equation}
Assuming analyticity of $V(x)$, this equation can be iteratively
solved for large $z$ and the solution which is square integrable
at $y\rightarrow\infty$ is given by
\begin{equation}\label{tam2}
    R(y,z)=\sqrt{y}K_\nu(y)+\sum_{n=0}^{\infty}\psi_n(y)z^{-1-n/2}\,,
\end{equation}
where $K_\nu(y)$ is the modified Bessel function and $\psi_n(y)$
depend polynomially on $V(x)$ and its derivatives. From this
expression it can be easily seen that the behavior of $R(y,z)$ for
$x\rightarrow 0^+$ is given by
\begin{equation}\label{venelori}
    R(y,z)\simeq\frac{\Gamma(\nu)}{2^{1-\nu}}\ y^{-\nu+1/2}+
    \frac{\Gamma(-\nu)}{2^{1+\nu}}\
    H(z)\cdot y^{\nu+1/2}
    +O(y^{3/2})
\end{equation}
where $H(z)$ admits a large-$z$ asymptotic expansion in
half-integer powers of $z$.

\bs

From eq.\ (\ref{solres}) for the case $\theta=infty$ and eqs.\
(\ref{Renelori}) and (\ref{venelori}) we obtain
\begin{equation}\label{kyh}
    K(z)=4^{\nu}\frac{\Gamma(1+\nu)}{\Gamma(1-\nu)}\;z^{-\nu}H(z)^{-1}\,.
\end{equation}
Since $H(z)$ admits an asymptotic series in half-integer powers of
$z$, the large-$z$ asymptotic expansion of $K(z)$ contains powers
of $z$ whose exponents depend on the parameter $\nu$. Proceeding
in a similar way one can prove, by means of expression
(\ref{tam2}), that ${\rm
Tr}\left\{(A^{0}+z)^{-1}-(A^{\infty}+z)^{-1}\right\}$ admits an
asymptotic expansion in half-integer powers of $z$.

\bs

From Theorem \ref{elthm} and due to the factor $z^{-\nu}$ in eq.\
(\ref{kyh}) we conclude that the large-$|\lambda|$ asymptotic
expansion of ${\rm
Tr}\left\{(A^{\theta}-\lambda)^{-1}-(A^{\infty}-\lambda)^{-1}\right\}$
contains integer powers of $\lambda^{-\nu}$. Finally, it can be
straightforwardly shown that its inverse Laplace transform, ${\rm
Tr}\left\{e^{-tA^\theta}-e^{-tA^\infty}\right\}$, admits the
asymptotic expansion given by (\ref{weshow}).

\bigskip

\noindent {\bf Acknowledgements:} The authors thank R.T.\ Seeley
for useful discussions, especially regarding the proof of Theorem
$2.1$. They also acknowledge support from Universidad Nacional de
La Plata (grant 11/X381) and CO\-NI\-CET (PIP 6160), Argentina.
P.A.G.P.\ would like to thank the participants of QFEXT'05, the
organizing comitee and, in particular, the warm hospitality of
Emilio Elizalde.

\appendix

\section{Proof of Theorem $2.1$}

By virtue of Riesz representation lemma \cite{Riesz}
\begin{equation}
\psi\in\mathcal{D}(A^{\dagger})\rightarrow
\exists\,\tilde{\psi}\in\mathbf{L}_2(\mathbb{R^+})\,:(\psi,A\phi)=(\tilde{\psi},\phi)\quad
\forall \phi\in\mathcal{D}(A)\,.
\end{equation}
Consequently,
\begin{equation}
A^{\dagger}\psi:=\tilde{\psi}\,.
\end{equation}
Defining $\chi:=x^{-\nu-1/2}\psi$ we obtain
\begin{equation}
    \partial_x(x^{2\nu+1}\partial_x\chi)=-x^{\nu+1/2}(\tilde{\psi}-V(x)\psi)
    \in\mathbf{L}_1(\mathbb{R^+})\,.
\end{equation}
Therefore, there exists a constant $C_1\in\mathbb{C}$ such that
\begin{equation}
    \partial_x\chi=C_1 x^{-1-2\nu}-x^{-1-2\nu}\int_0^x
    y^{\nu+1/2}\left(-\partial_y^2+\frac{\nu^2-1/4}{y^2}\right)\psi\,dy\,.
\end{equation}
Cauchy-Schwartz inequality implies
\begin{eqnarray}
    \left|\,x^{-1-2\nu}\int_0^x y^{\nu+1/2}\left(-\partial_y^2+
    \frac{\nu^2-1/4}{y^2}\right)\psi\,dy\,\right|\leq\nonumber\\
    C_2\,\left\|\left(-\partial_y^2+
    \frac{\nu^2-1/4}{y^2}\right)\psi\right\|_{(0,x)}\,x^{-\nu}\,,
\end{eqnarray}
for some $C_2\in\mathbb{C}$. In consequence,
\begin{eqnarray}
    \left|\,\int^x z^{-1-2\nu}\int_0^z
    y^{\nu+1/2}\left(-\partial_x^2+
    \frac{\nu^2-1/4}{x^2}\right)\psi\,dy\,dz\,\right|\leq\nonumber\\
    C_3+C_4\,x^{1-\nu}\,,
\end{eqnarray}
where $C_3,C_4\in\mathbb{C}$. Thus, there exist
$C_5,C_6\in\mathbb{C}$, such that
\begin{equation}
    \psi=C_5\,x^{-\nu+1/2}+C_6\,x^{\nu+1/2}+O(x^{3/2})\,,
\end{equation}
for $x\rightarrow 0^+$.\begin{flushright}$\Box$\end{flushright}


\end{document}